\documentclass[conference]{IEEEtran}
\IEEEoverridecommandlockouts

\usepackage{cite}
\usepackage{algpseudocode}
\usepackage{graphicx}
\usepackage{amsmath}
\usepackage{amsfonts}
\usepackage{epstopdf}
\usepackage{xcolor}
\usepackage{enumerate}
\usepackage{algorithm}
\usepackage{multirow}
\usepackage{transparent}
\usepackage{subfigure}
\usepackage{float}

\def\BibTeX{{\rm B\kern-.05em{\sc i\kern-.025em b}\kern-.08em
    T\kern-.1667em\lower.7ex\hbox{E}\kern-.125emX}}

\newtheorem{corollary}{Corollary}
\newtheorem{theorem}{Theorem}

\newtheorem{lemma}{Lemma}

\newcommand{\T}{\textnormal{T}}
\newcommand{\R}{\textnormal{R}}
\let\ss= \scriptscriptstyle
\begin{document}
		\title{Membrane Fusion-Based Transmitter Design for Molecular Communication Systems}
\author{\IEEEauthorblockN{Xinyu Huang\IEEEauthorrefmark{1}, Yuting Fang\IEEEauthorrefmark{2}, Adam Noel\IEEEauthorrefmark{3}, and Nan Yang\IEEEauthorrefmark{1}}
	\IEEEauthorblockA{\IEEEauthorrefmark{1}School of Engineering, Australian National University, Canberra, ACT, Australia}
	\IEEEauthorblockA{\IEEEauthorrefmark{2}Department of Electrical and Electronic Engineering, University of Melbourne, Parkville, VIC, Australia}
	\IEEEauthorblockA{\IEEEauthorrefmark{3}School of Engineering, University of Warwick, Coventry, CV4 7AL, UK}
	\IEEEauthorblockA{Email: xinyu.huang1@anu.edu.au, yuting.fang@unimelb.edu.au, adam.noel@warwick.ac.uk, nan.yang@anu.edu.au}}

\maketitle

\begin{abstract}
This paper proposes a novel imperfect spherical transmitter (TX) model, namely the membrane fusion (MF)-based TX, that adopts MF between a vesicle and the TX membrane to release molecules encapsulated within the vesicle. For the MF-based TX, the molecule release probability and the fraction of molecules released from the TX membrane are derived. Incorporating molecular degradation and a fully-absorbing receiver (RX), the end-to-end molecule hitting probability at the RX is also derived. A simulation framework for the MF-based TX is proposed, where the released point on the TX membrane and the released time of each molecule are determined. Aided by the simulation framework, the derived analytical expressions are validated. Simulation results verify that a low MF probability or low vesicle mobility slows the release of molecules from the TX, extends time required to reach the peak release probability, and reduces the end-to-end molecule hitting probability at the RX.
\end{abstract}


\section{Introduction}

Molecular communication (MC) has become one of the most promising methods for nanoscale communication. In MC, molecules act as information carriers. The main driving force behind engineering MC is its diversity of potential applications in the medical field, e.g., lab-on-a-chip devices, cell-on-chip devices, and targeted drug delivery \cite{farsad2016comprehensive}. An end-to-end MC channel model incorporates a transmitter $(\mathrm{TX})$, the propagation environment, and a receiver $(\mathrm{RX})$. Molecule propagation environment and reception mechanism at the $\mathrm{RX}$ have been widely investigated in previous studies, e.g., \cite{ahmadzadeh2016comprehensive,jamali2019channel}. However, few studies have investigated the impact of signaling pathways inside the $\mathrm{TX}$ and the interaction of molecular signals with the $\mathrm{TX}$ surface on the MC system performance.

Most existing studies assumed the $\mathrm{TX}$ to be an ideal point source that can release molecules instantaneously \cite{jamali2019channel}. Compared to realistic scenarios, this ideal $\mathrm{TX}$ neglects the effects of $\mathrm{TX}$ geometry, signaling pathways inside the $\mathrm{TX}$, and chemical reactions during the release process. Recently, some studies considered these effects of the $\mathrm{TX}$, e.g., \cite{yilmaz2017chemical,arjmandi2016ion,schafer2019spherical}.\cite{yilmaz2017chemical} proposed a spherical $\mathrm{TX}$ that reflects the emitted molecules and investigated the directivity gain achieved by the reflecting $\mathrm{TX}$.  \cite{arjmandi2016ion} proposed an ion channel based $\mathrm{TX}$, where molecule release is controlled by opening and closing ion channels. \cite{schafer2019spherical} considered a spherical $\mathrm{TX}$ with a semi-permeable boundary whose permeability is used to control molecule release. Although these studies stand on their own merits, none of them has considered an exocytosis-like mechanism for molecule release.

In nature, exocytosis is a form of active transport in which a cell transports molecules out of the cell by secreting them through an energy-dependent process \cite{jahn1999membrane}. Exocytosis is common for cells because many chemical substances are large molecules and cannot pass through the cell membrane by passive means \cite{biomor}. In exocytosis, vesicles\footnote{A vesicle is a small, round or oval-shaped container for the storage of molecules, and as compartments with particular chemical reactions \cite{bonifacino2004mechanisms}.} are carried to the cell membrane to secrete their contents into the extracellular environment. This secretion is performed by the membrane fusion (MF) that fuses the vesicle with the cell membrane. When the vesicle moves close to the cell membrane, the v-SNARE protein on the vesicle membrane binds to the t-SNARE protein on the cell membrane to generate the \textit{trans}-SNARE complex that catalyzes MF \cite{bonifacino2004mechanisms}. Unlike the ideal point $\mathrm{TX}$, exocytosis constraints the release of molecules. Hence, a $\mathrm{TX}$ that uses MF to release molecules merits investigation.

In this paper, we propose a novel $\mathrm{TX}$ model in a three-dimensional (3D) environment, namely the MF-based $\mathrm{TX}$, which uses fusion between a vesicle generated within the $\mathrm{TX}$ and the $\mathrm{TX}$ membrane to release molecules encapsulated within the vesicle. By considering a fully-absorbing $\mathrm{RX}$ that absorbs molecules once they hit the $\mathrm{RX}$ surface, we investigate the end-to-end channel impulse response (CIR) between the MF-based $\mathrm{TX}$ and the $\mathrm{RX}$, where the CIR is the hitting probability of molecules at the RX \cite{jamali2019channel}.

Our major contributions are summarized as follows. We first derive the time-varying molecule release probability and the fraction of molecules released from the $\mathrm{TX}$ by a given time. We then derive the end-to-end molecule hitting probability at the $\mathrm{RX}$ due to the MF-based $\mathrm{TX}$. Furthermore, we propose a simulation framework for the MF-based $\mathrm{TX}$ model to simulate the diffusion and fusion of vesicles within the $\mathrm{TX}$. In this simulation framework, the release point on the $\mathrm{TX}$ membrane and the release time of each molecule are determined. Aided by the proposed simulation framework, we demonstrate the accuracy of our analytical derivations. Our numerical results show that a low MF probability or low vesicle mobility slows the release of molecules from the $\mathrm{TX}$, increases the time to reach the peak release probability, and reduces the end-to-end molecule hitting probability at the $\mathrm{RX}$.
\section{System Model}\label{sm}
\begin{figure}[!t]
	\begin{center}
		\includegraphics[height=1.5in,width=0.8\columnwidth]{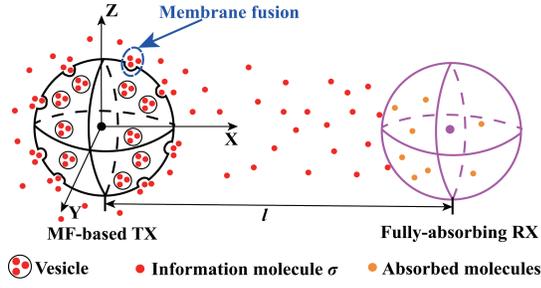}
		\caption{Illustration of the system model, where one MF-based $\mathrm{TX}$ communicates with one fully-absorbing $\mathrm{RX}$ in a 3D environment.}\label{sys}\vspace{-0.5em}
	\end{center}
	\vspace{-4mm}
\end{figure}
In this paper, we consider an unbounded 3D environment, where an MF-based $\mathrm{TX}$ communicates with a fully-absorbing $\mathrm{RX}$, as depicted in Fig. \ref{sys}. Both $\mathrm{TX}$ and $\mathrm{RX}$ are spheres with radius $r_{\ss\T}$ and $r_{\ss\R}$, respectively. The center of the $\mathrm{TX}$ is chosen as the origin of the environment. The center of the $\mathrm{RX}$ is distance $l$ away from the center of the $\mathrm{TX}$. We assume that the spherical $\mathrm{TX}$ releases molecules from its outer membrane after fusion between the membrane and vesicles. Each vesicle stores and transports $\eta$ molecules of type $\sigma$. We consider that the $\mathrm{TX}$ is filled with a fluid medium that has uniform temperature and viscosity. We also consider an impulse of $N_\mathrm{v}$ vesicles released within the $\mathrm{TX}$ at $t=0$. The experiment in \cite{kyoung2008vesicle} demonstrates that vesicles diffuse in a fluid medium with a diffusion coefficient. Based on \cite{kyoung2008vesicle}, we assume that once vesicles are released, they diffuse randomly with a constant diffusion coefficient $D_\mathrm{v}$. According to \cite{bonifacino2004mechanisms}, natural fusion of a vesicle and the cell membrane can be considered as two steps: 1) The v-SNARE protein ($\mathrm{S}_\mathrm{v}$)  on the vesicle membrane binds to the t-SNARE protein ($\mathrm{S}_\mathrm{t}$) on the cell membrane to generate the \textit{trans}-SNARE complex ($\mathrm{S}_\mathrm{c}$), and 2) $\mathrm{S}_\mathrm{c}$ catalyzes the fusion of vesicular and cell membranes. For tractability, we make the following assumptions on the $\mathrm{TX}$ model:
\begin{itemize}
	\item[A1)] Vesicles are released from the $\mathrm{TX}$'s center. This assumption simplifies the theoretical analysis since the $\mathrm{TX}$ model is symmetrical. Considering vesicles released from any point within the $\mathrm{TX}$ is an interesting future work.
	\item[A2)] The binding between $\mathrm{S}_\mathrm{v}$ and $\mathrm{S}_\mathrm{t}$ is modeled as an irreversible reaction, given by
	\begin{align}\label{IR}
	\mathrm{S}_\mathrm{v}+\mathrm{S}_\mathrm{t}\stackrel{k_\mathrm{f}}{\longrightarrow}\mathrm{S}_\mathrm{c},
	\end{align}
	where $k_\mathrm{f}$ is the forward reaction rate in $\mu\mathrm{m}/\mathrm{s}$. We acknowledge that the MF process in nature is more complex than this assumption. For instance, MF also depends on an increased intracellular calcium ($\mathrm{Ca}^{2+}$) concentration \cite{hay2007calcium}. The irreversible reaction modeled in \eqref{IR} is a good first step to incorporate the MF mechanism in MC modeling.
	\item[A3)] The membrane is fully covered by an infinite number of $\mathrm{S}_\mathrm{t}$ and the occupancy of $\mathrm{S}_\mathrm{t}$ is ignored. Assuming perfect receptor coverage and ignoring occupancy are for tractability and
	have been adopted in several previous studies, e.g., \cite{arjmandi2019diffusive,ahmadzadeh2016comprehensive}.
	\item[A4)] The generation of $\mathrm{S}_\mathrm{c}$ guarantees MF. In nature, $\mathrm{S}_\mathrm{c}$ catalyzes the MF process. Hence, this assumption becomes reasonable if the reaction rate of this catalytic reaction is assumed to be infinity.
	\item[A5)] Once molecules are released, the spherical $\mathrm{TX}$ does not hinder the random diffusion of molecules in the propagation environment, i.e, the $\mathrm{TX}$ is transparent to the diffusion of released molecules. \cite{yilmaz2017chemical} analyzed the hindrance of the $\mathrm{TX}$ membrane to the diffusion of molecules using simulation. Considering both the hindrance of the $\mathrm{TX}$ membrane and the absorbing $\mathrm{RX}$ is cumbersome for theoretical analysis. We will investigate the impact of the $\mathrm{TX}$ membrane on molecule diffusion in the propagation environment in future work.
\end{itemize}
Based on A2-A4, if a vesicle hits the $\mathrm{TX}$ membrane, it fuses to the membrane with a probability of $k_\mathrm{f}\sqrt{\frac{\pi\Delta t}{D_\mathrm{v}}}$ during a time interval $\Delta t$ \cite{erban2007reactive}. We define this probability as the MF probability. After MF, the molecules $\sigma$ stored by the vesicle are released into the propagation environment. The location and time for the occurrence of MF are the initial location and time for molecules to start moving in the propagation environment.

We assume that the propagation environment outside the spherical $\mathrm{TX}$ and $\mathrm{RX}$ is a fluid medium with uniform temperature and viscosity. Once information molecules $\sigma$ are released from the $\mathrm{TX}$, they diffuse randomly with a constant diffusion coefficient $D_{\sigma}$. Moreover, we consider unimolecular degradation in the propagation environment, where type $\sigma$ molecules can degrade into some other molecular species $\hat{\sigma}$ that cannot be identified by the $\mathrm{RX}$, i.e., $\sigma\stackrel{k_\mathrm{d}}{\longrightarrow}\hat{\sigma}$ \cite[Ch. 9]{chang2005physical}, where $k_\mathrm{d}\; \left[\mathrm{s}^{-1}\right]$ is the degradation rate. In addition, we model the $\mathrm{RX}$ as a spherical fully-absorbing $\mathrm{RX}$. Molecules $\sigma$ are absorbed as soon as they hit the $\mathrm{RX}$ surface.
\section{Derivation of Channel Impulse Response}\label{CIR}
In this section, we first derive the molecule release probability and the fraction of released molecules from the $\mathrm{TX}$ membrane due to the impulsive emission of vesicles from the $\mathrm{TX}$'s center. We define the molecule release probability as the probability of one molecule being released at time $t$ from the $\mathrm{TX}$ membrane, when this molecule is released from the origin at time $t=0$. We then derive the molecule hitting probability at the $\mathrm{RX}$ when the $\mathrm{TX}$ releases molecules uniformly over the $\mathrm{TX}$ membrane. We define the molecule hitting probability as the probability of one molecule hitting the $\mathrm{RX}$ at time $t$ when this molecule is released at time $t=0$. Using the previously derived release probability and hitting probability, we finally derive the end-to-end molecule hitting probability at the $\mathrm{RX}$ due to the impulsive emission of vesicles from the $\mathrm{TX}$'s center.
\subsection{Release Probability from TX Membrane}\label{rr}
As each molecule is released when MF occurs, the molecule release probability equals the fusion probability of vesicles. Thus, we need to obtain the distribution function of vesicles within the $\mathrm{TX}$ to derive the molecule release probability from the $\mathrm{TX}$ membrane. In the spherical coordinate system, we denote $C(r,t)$, $0\leq r\leq r_{\ss\T}$, as the vesicle distribution function at time $t$ with distance $r$ from the $\mathrm{TX}$'s center. When an impulse of vesicles is released from the $\mathrm{TX}$'s center at $t=0$, the initial condition is expressed as \cite[eq. 3(c)]{dincc2018impulse}
\begin{align}\label{IC}
C(r,t\rightarrow 0)=\frac{1}{4\pi r^2}\delta(r),
\end{align}
where $\delta(\cdot)$ is the Dirac delta function. According to Fick's second law, the diffusion of vesicles inside the $\mathrm{TX}$ can be described as \cite{berg1993random}
\begin{align}\label{fsl}
D_\mathrm{v}\frac{\partial^2\left(rC(r,t)\right)}{\partial r^2}=\frac{\partial \left(rC(r,t)\right)}{\partial t}.
\end{align}

Based on A2 and A4, the boundary condition is described by the irreversible reaction given by \eqref{IR}, which can be characterized by the third type (Robin) boundary condition as \cite{crank1979mathematics}
\begin{align}\label{bc}
D_\mathrm{v}\frac{\partial C(r,t)}{\partial r}\big|_{r=r_{\ss\T}}=-k_\mathrm{f}C(r_{\ss\T},t),
\end{align}
where the negative sign on the right-hand side indicates that the condition is over the inner boundary.

Based on the initial condition in \eqref{IC}, Fick's second law in \eqref{fsl}, and the boundary condition in \eqref{bc}, we derive the closed-form expression for the release probability from the $\mathrm{TX}$ membrane, denoted by $f_\mathrm{r}(t)$, in the following theorem:
\begin{theorem}\label{theorem1}
	The release probability of molecules from the $\mathrm{TX}$ membrane at time $t$ is given by
	\begin{align}\label{ff}
	f_\mathrm{r}(t)=\sum_{n=1}^{\infty}\frac{4r_{\ss\T}^2k_\mathrm{f}\lambda_n^3}{2\lambda_nr_{\ss\T}-\mathrm{sin}\left(2\lambda_nr_{\ss\T}\right)}j_0(\lambda_nr_{\ss\T})\exp\left(-D_\mathrm{v}\lambda_n^2t\right),
	\end{align}
	where $j_0(\cdot)$ is the zeroth order of the first type of the spherical Bessel function \cite{olver1960bessel},  $n=1,2,3,...$, and $\lambda_n$ is obtained by solving
	\begin{align}\label{Dvl}
	D_\mathrm{v}\lambda_nj_0^{'}\left(\lambda_nr_{\ss\T}\right)=-k_\mathrm{f}j_0\left(\lambda_nr_{\ss\T}\right),
	\end{align}
where $j_0^{'}(z)=\frac{\partial j_0(z)}{\partial z}$.
\end{theorem}
\begin{IEEEproof}
	Please see Appendix \ref{app}.
\end{IEEEproof}

 We denote $F_\mathrm{r}(t)$ as the fraction of molecules released by time $t$ and obtain it by $F_\mathrm{r}(t)=\int_{0}^{t}f_\mathrm{r}(u)\mathrm{d}u$. We present $F_\mathrm{r}(t)$ in the following corollary:
\begin{corollary}
	The fraction of released molecules from the $\mathrm{TX}$ by time $t$ is given by
	\begin{align}\label{num}
	F_\mathrm{r}(t)=\sum_{n=1}^{\infty}\frac{4r_{\ss\T}^2 k_\mathrm{f}\lambda_nj_0(\lambda_nr_{\ss\T})}{D_\mathrm{v}\left(2\lambda_nr_{\ss\T}-\mathrm{sin}(2\lambda_nr_{\ss\T})\right)}\left(1-\exp\left(-D_\mathrm{v}\lambda_n^2t\right)\!\right).
	\end{align}
\end{corollary}
The number of molecules released by time $t$ is $N_\mathrm{v}\eta F_\mathrm{r}(t)$.
\subsection{Hitting Probability at RX with Uniform Release of Molecules}\label{thr}
The aim of this paper is to derive the end-to-end hitting probability of molecules at the $\mathrm{RX}$ surface when an impulse of vesicles is released from the $\mathrm{TX}$'s center. To this end, we first derive the hitting probability when the molecules are uniformly released from the $\mathrm{TX}$ membrane, i.e., ignoring the internal molecules' propagation within the $\mathrm{TX}$ and the $\mathrm{TX}$ MF process. We consider the scenario that molecules are initially uniformly distributed over the $\mathrm{TX}$ membrane and released simultaneously at $t=0$, where the membrane area is denoted by $\Omega_{\ss\T}$. We denote $p_\mathrm{u}(t)$ as the corresponding hitting probability at the $\mathrm{RX}$ due to the uniform release of molecules over the $\mathrm{TX}$ membrane. We note that uniformly-distributed molecules means that the likelihood of a molecule released from any point on the $\mathrm{TX}$ membrane is the same. We denote this probability by $\rho$, where we have $\rho=\left(4\pi r_{\ss\T}^2\right)^{-1}$. We further consider an arbitrary point $\alpha$ on the $\mathrm{TX}$ membrane. Based on \cite[eq. (9)]{heren2015effect}, the hitting probability $p_\alpha(t)$ of a molecule at the $\mathrm{RX}$ at time $t$ when the molecule is released from the point $\alpha$ at time $t=0$ is given by
\begin{align}\label{pa}
p_\alpha(t)=\frac{r_{\ss\R}(l_\alpha-r_{\ss\R})}{l_\alpha\sqrt{4\pi D_{\sigma}t^3}}\exp\left(-\frac{\left(l_\alpha-r_{\ss\R}\right)^2}{4D_{\sigma}t}-k_\mathrm{d}t\right),
\end{align}
where $l_\alpha$ is the distance between the point $\alpha$ and the center of the $\mathrm{RX}$.

Given that molecules are distributed uniformly over the $\mathrm{TX}$ membrane, $p_\mathrm{u}(t)$ is obtained by taking the surface integral of $p_\alpha(t)$ over the spherical $\mathrm{TX}$ membrane. Using this method, we solve $p_\mathrm{u}(t)$ in the following lemma:
\begin{lemma}\label{hrb}
	The hitting probability of molecules at the $\mathrm{RX}$ at time $t$ when the $\mathrm{TX}$ uniformly releases the molecules over the $\mathrm{TX}$ membrane at time $t=0$ is given by
	\begin{align}\label{pbt}
	p_\mathrm{u}(t)\!\!=\!\!\frac{2\rho r_{\ss\T}r_{\ss\R}}{l}\!\sqrt{\frac{\pi D_{\sigma}}{t}}\!\left[\!\exp\!\left(\!\!-\frac{\beta_1}{t}\!-\!k_\mathrm{d}t\!\right)\!-\!\exp\!\left(\!-\frac{\beta_2}{t}\!-\!k_\mathrm{d}t\right)\!\right],
	\end{align}
	where $\beta_1=\frac{(r_{\ss\T}+r_{\ss\R})(r_{\ss\T}+r_{\ss\R}-2l)+l^2}{4D_{\sigma}}$ and $\beta_2=\frac{(r_{\ss\T}-r_{\ss\R})(r_{\ss\T}-r_{\ss\R}+2l)+l^2}{4D_{\sigma}}$.
\end{lemma}
\begin{IEEEproof}
	Please see Appendix \ref{ab}.
\end{IEEEproof}

\subsection{End-to-End Hitting Probability at the RX}
We denote $p_\mathrm{v}(t)$ as the end-to-end hitting probability of molecules at the $\mathrm{RX}$ when an impulse of vesicles is released from the $\mathrm{TX}$'s center at time $t=0$. The release probability of molecules from the $\mathrm{TX}$ at time $u$, $0\leq u\leq t$, is given by \eqref{ff}, which is $f_\mathrm{r}(u)$. For molecules released at time $u$, the hitting probability of molecules at the $\mathrm{RX}$ at time $t$ is given by \eqref{pbt}, which is $p_\mathrm{u}(t-u)$. Based on that, $p_\mathrm{v}(t)$ is given by
\begin{align}\label{conv}
	p_\mathrm{v}(t)=\int_{0}^{t}p_\mathrm{u}(t-u)f_\mathrm{r}(u)\mathrm{d}u.
\end{align}
Substituting \eqref{ff} and \eqref{pbt} into \eqref{conv}, we derive $p_\mathrm{v}(t)$ in the following theorem:
\begin{theorem}
	The end-to-end hitting probability of molecules at the $\mathrm{RX}$ at time $t$ for an impulsive emission of vesicles from the $\mathrm{TX}$'s center at time $t=0$, is given by
	\begin{align}\label{pvt}
	p_\mathrm{v}(t)=&\frac{8\rho r_{\ss\T}^3r_{\ss\R}k_\mathrm{f}\sqrt{\pi D_{\sigma}}\exp\left(-k_\mathrm{d}t\right)}{l}\sum_{n=1}^{\infty}\frac{\lambda_n^3j_0\left(\lambda_nr_{\ss\T}\right)}{2\lambda_nr_{\ss\T}-\mathrm{sin}\left(2\lambda_nr_{\ss\T}\right)}\notag\\&\times\left[\varepsilon(\beta_1,t)-\varepsilon(\beta_2,t)\right],
	\end{align}
	where
	\begin{align}
		\varepsilon(\zeta,t)\!\!=\!\!\!\int_{0}^{t}\!\!\frac{1}{\sqrt{t\!-\!u}}\exp\left(-\frac{\zeta}{t-u}-(D_\mathrm{v}\lambda_n^2-k_\mathrm{d})u\right)\mathrm{d}u.
	\end{align}
$\varepsilon(\zeta,t)$ can be calculated numerically using the MATLAB.
\end{theorem}

\section{Simulation Framework for MF-based TX}
In this section, we describe the stochastic simulation framework for the MF-based $\mathrm{TX}$. We use a particle-based simulation method that records the exact position of each vesicle. For simulating molecules' diffusion in the propagation environment and absorption by the $\mathrm{RX}$, the particle-based simulation is also applied, which is omitted here due to the page limit.
\subsection{Emission and Diffusion}
$N_\mathrm{v}$ vesicles are released from the $\mathrm{TX}$'s center at $t=0$. We denote $\Delta t_\mathrm{s}$ as the simulation interval. After vesicles are released, they perform Brownian motion during each simulation interval. Therefore, the displacements of each vesicle in three dimensions during the simulation interval are independent Gaussion random variables (RVs) with zero mean and variance $2D_\mathrm{v}\Delta t_\mathrm{s}$.
\subsection{Fusion or Reflection}
We denote the locations of a vesicle at the start and end of the $\gamma$th simulation interval by $(x_{\gamma-1}, y_{\gamma-1}, z_{\gamma-1})$ and $(x_{\gamma}, y_\gamma, z_\gamma)$, respectively. If the distance between a vesicle and the $\mathrm{TX}$'s center is larger than $r_{\ss\T}$ at the end of the $\gamma$th interval, we assume that the vesicle has hit the $\mathrm{TX}$ membrane. As described in Section \ref{sm}, this vesicle then fuses with the $\mathrm{TX}$ membrane with probability $k_\mathrm{f}\sqrt{\frac{\pi\Delta t_\mathrm{s}}{D_\mathrm{v}}}$ and is reflected with probability $1-k_\mathrm{f}\sqrt{\frac{\pi\Delta t_\mathrm{s}}{D_\mathrm{v}}}$.

Molecules stored in a vesicle are released at the time and location where the vesicle fused with the $\mathrm{TX}$ membrane. Thus, we need to derive where and when fusion with the membrane occurred. For a vesicle fusing with the $\mathrm{TX}$ membrane during the $\gamma$th simulation interval, we assume that the intersection between the line that is formed by $(x_{\gamma-1}, y_{\gamma-1}, z_{\gamma-1})$ and $(x_\gamma, y_\gamma, z_\gamma)$ and the $\mathrm{TX}$ membrane is the fusion point whose coordinates, denoted by $(x_{\mathrm{f}, \gamma}, y_{\mathrm{f}, \gamma}, z_{\mathrm{f}, \gamma})$, are given by \cite{noel2017simulating}
\begin{align}\label{xr}
x_{\mathrm{f}, \gamma}=\frac{-\Lambda_2\pm\sqrt{\Lambda_2^2-4\Lambda_1\Lambda_3}}{2\Lambda_1},
\end{align}
\begin{align}\label{yr}
y_{\mathrm{f}, \gamma}=\frac{(x_{\mathrm{f}, \gamma}-x_{\gamma-1})(y_\gamma-y_{\gamma-1})}{x_\gamma-x_{\gamma-1}}+y_{\gamma-1},
\end{align}
and
\begin{align}\label{zr}
z_{\mathrm{f}, \gamma}=\frac{(x_{\mathrm{f}, \gamma}-x_{\gamma-1})(z_\gamma-z_{\gamma-1})}{x_\gamma-x_{\gamma-1}}+z_{\gamma-1},
\end{align}
respectively, where $\Lambda_1=(x_\gamma-x_{\gamma-1})^2+(y_\gamma-y_{\gamma-1})^2+(z_\gamma-z_{\gamma-1})^2$, $\Lambda_2=2(y_\gamma-y_{\gamma-1})(x_\gamma y_{\gamma-1}-x_{\gamma-1}y_\gamma)+2(z_\gamma-z_{\gamma-1})(x_\gamma z_{\gamma-1}-x_{\gamma-1}z_\gamma)$, and $\Lambda_3=x_{\gamma-1}(y_\gamma-y_{\gamma-1})(x_{\gamma-1}y_\gamma-2y_{\gamma-1}x_\gamma+y_{\gamma-1}x_\gamma)+x_{\gamma-1}(z_\gamma-z_{\gamma-1})(x_{\gamma-1}z_\gamma-2x_\gamma z_{\gamma-1}+x_{\gamma-1}z_{\gamma-1})+(x_\gamma-x_{\gamma-1})^2(y_{\gamma-1}^2+z_{\gamma-1}^2-r_{\ss\T})$. In \eqref{xr}, $x_{\mathrm{f}, \gamma}$ is chosen by satisfying $(x_{\mathrm{f}, \gamma}-x_{\gamma-1})(x_{\mathrm{f}, \gamma}-x_\gamma)<0$. We denote $t_{\gamma-1}$ as the start time of the $\gamma$th simulation interval and $t_{\gamma-1}+\Delta t_{\mathrm{f}, \gamma}$ as the fusion time of the vesicle in the $\gamma$th simulation interval. As each vesicle follows Brownian motion, the square of the displacement is proportional to the time within the same simulation interval. Therefore, we derive $\Delta t_{\mathrm{f}, \gamma}$ as
\begin{align}\label{tr}
\Delta t_{\mathrm{f}, \gamma}\!\!=\!\!\frac{(x_{\mathrm{f}, \gamma}\!-\!x_{\gamma-1})^2+(y_{\mathrm{f},\gamma}\!-\!y_{\gamma-1})^2+(z_{\mathrm{f},\gamma}\!-\!z_{\gamma-1})^2}{(x_\gamma-x_{\gamma-1})^2+(y_\gamma-y_{\gamma-1})^2+(z_\gamma-z_{\gamma-1})^2}\Delta t_\mathrm{s}.
\end{align}

For vesicles failing to fuse with the $\mathrm{TX}$ membrane, we make the assumption that they are sent back to their positions at the start of the current simulation interval.
\section{Numerical Results and Discussions}
\begin{figure}[!t]
	\centering
	
	\subfigure[Releasing probability]{
		\begin{minipage}[t]{1\linewidth}
			\centering
			\includegraphics[width=2.7in]{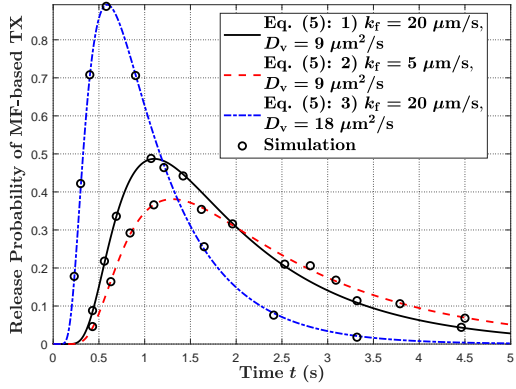}
			\label{a}
		\end{minipage}%
	}%
	\quad
	\subfigure[The number of molecules released]{
		\begin{minipage}[t]{0.9\linewidth}
			\centering
			\includegraphics[width=2.7in]{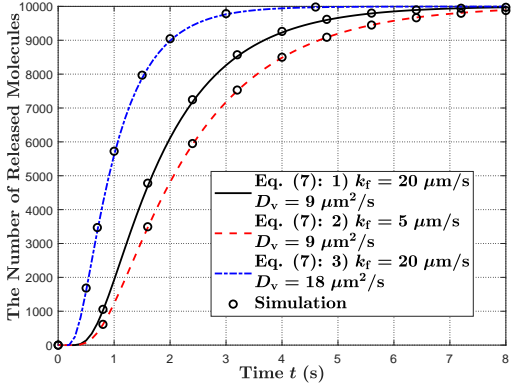}
			\label{b}
		\end{minipage}%
	}
	\centering
	\caption{Molecule release probability from the MF-based $\mathrm{TX}$ at time $t$ and the number of molecules released from the MF-based $\mathrm{TX}$ by time $t$ versus time $t$ for three parameter sets.}\label{term}
\end{figure}

In this section, we present numerical results to validate our theoretical analysis and enable insightful discussion. The simulation time interval is $\Delta t_\mathrm{s}=0.001\;\mathrm{s}$ and all results are averaged over 5000 realizations. Throughout this section, we set $N_\mathrm{v}=100$, $\eta=100$, $r_{\ss\T}=r_{\ss\R}=10\;\mu\mathrm{m}$, $D_\mathrm{v}=9\;\mu\mathrm{m}^2/\mathrm{s}$, $k_\mathrm{f}=20\;\mu\mathrm{m}/\mathrm{s}$, $l=40\;\mu\mathrm{m}$, $k_\mathrm{d}=0.8\;\mathrm{s}^{-1}$, and $D_{\sigma}=1000\;\mu\mathrm{m}^2/\mathrm{s}$ \cite{kyoung2008vesicle,arjmandi2019diffusive}, unless otherwise stated. In Figs. 2-4, we vary the forward reaction rate and vesicle diffusion coefficient to investigate their impact on molecule release, time to reach the peak release probability, and molecule absorption. In Fig. \ref{term} and Fig. \ref{hr}, we observe precise agreement between our simulation results and the analytical curves generated from Section \ref{CIR}, which demonstrate the accuracy of our analysis.

In Fig. \ref{term}, we plot the molecule release probability from the $\mathrm{TX}$ at time $t$ versus time $t$ in Fig. 2(a) and the number of released molecules from the $\mathrm{TX}$ until time $t$ versus time $t$ in Fig. 2(b). First, by comparing parameter sets 1) and 2) in Fig. 2(a), we observe that the peak release probability decreases and the tail of the release probability becomes longer with a decrease in $k_\mathrm{f}$. This is because the decrease in $k_\mathrm{f}$ reduces the fusion probability between a vesicle and the $\mathrm{TX}$ membrane. Second, by comparing parameter sets 1) and 3) in Fig. 2(a), we observe that the peak release probability decreases and a longer time is required for the $\mathrm{TX}$ to start releasing molecules with a decrease in $D_\mathrm{v}$. This is because decreasing $D_\mathrm{v}$ slows down the vesicle. Third, in Fig. 2(b), we observe that all molecules are released from the $\mathrm{TX}$ with sufficient long time due to the impulsive release.

\begin{figure}[!t]
	\begin{center}
		\includegraphics[height=2in,width=0.8\columnwidth]{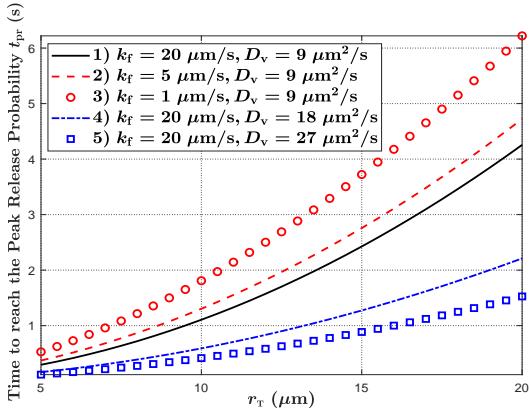}
		\caption{Time to reach the peak molecule release probability from the $\mathrm{TX}$ versus $r_{\ss\T}$ for five parameter sets.}\label{hig}\vspace{-0.5em}
	\end{center}
	\vspace{-4mm}
\end{figure}

In Fig. \ref{hig}, we plot the time to reach the peak molecule release probability from the $\mathrm{TX}$, denoted by $t_\mathrm{pr}$, versus the radius of the $\mathrm{TX}$ by searching the highest value in \eqref{ff} and recording the corresponding time. First, we observe that the time to reach the peak release probability increases with an increase in $r_{\ss\T}$. This is because a vesicle needs to diffuse for longer to arrive at the $\mathrm{TX}$ membrane with a larger radius. Second, by comparing parameter sets 1), 2), and 3), we observe that the time increases with a decrease in $k_\mathrm{f}$. This is because lower $k_\mathrm{f}$ reduces the fusion probability such that it takes longer to reach the peak release probability. Third, by comparing parameter sets 1), 4), and 5), we observe that the time increases with a decrease in $D_\mathrm{v}$. This is because the smaller value of $D_\mathrm{v}$ slows down the vesicle such that a longer time is required for the vesicle to arrive at the $\mathrm{TX}$ membrane.

\begin{figure}[!t]
	\begin{center}
		\includegraphics[height=2in,width=0.8\columnwidth]{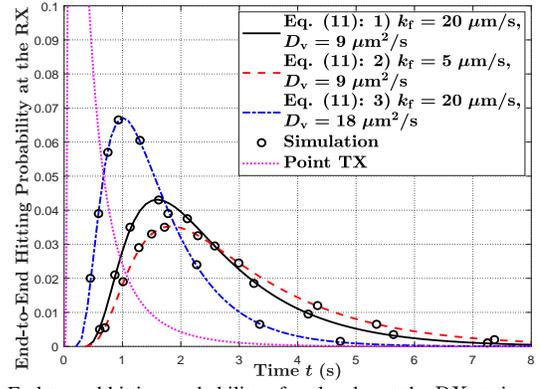}\vspace{-1em}
		\caption{End-to-end hitting probability of molecules at the $\mathrm{RX}$ at time $t$ versus time $t$ for three parameter sets.}\label{hr}\vspace{-1.5em}
	\end{center}
\end{figure}

In Fig. \ref{hr}, we plot the end-to-end hitting probability of molecules at the $\mathrm{RX}$ at time $t$ versus time $t$. First, by comparing parameter sets 1) and 2), we observe that the peak end-to-end hitting probability decreases with a decrease in $k_\mathrm{f}$. This is because decreasing $k_\mathrm{f}$ reduces the release probability of molecules such that the end-to-end hitting probability at the $\mathrm{RX}$ decreases. Second, by comparing parameter sets 1) and 3), we observe that the peak end-to-end hitting probability increases and less time is required for molecules to start hitting the $\mathrm{RX}$ with an increase in $D_\mathrm{v}$. This is because the larger value of $D_\mathrm{v}$ enables the earlier release of molecules and increases the peak release probability. Third, to show the difference between the MF-based $\mathrm{TX}$ and an ideal point $\mathrm{TX}$, we also plot the hitting probability of an ideal point $\mathrm{TX}$ based on \cite[eq. (9)]{heren2015effect}. We observe that the peak value of the hitting probability curve for the point $\mathrm{TX}$ is larger than that for the MF-based $\mathrm{TX}$, i.e., the hitting probability curve for the point $\mathrm{TX}$ is narrower. This is because molecules are instantaneously released from the point $\mathrm{TX}$ while molecules need to be transported to $\mathrm{TX}$ membrane before release.

\section{Conclusion}
In this paper, we proposed a new $\mathrm{TX}$ model that are based on MF between vesicles generated within the $\mathrm{TX}$ and the $\mathrm{TX}$ membrane to release molecules. We derived the molecule release probability and the fraction of released molecules from the $\mathrm{TX}$. By considering a fully-absorbing $\mathrm{RX}$ in the environment, we derived new closed-form expressions for the end-to-end molecule hitting probability at the $\mathrm{RX}$. We further propose a simulation framework for the MF-based $\mathrm{TX}$. Our results showed that our analytical expressions are accurate. They also showed that a low MF probability or low vesicle mobility slows the release of molecules, extends the time ro reach the peak release probability, and reduces the end-to-end molecule hitting probability at the $\mathrm{RX}$. Future work includes 1) considering the reflecting membrane of the $\mathrm{TX}$ and 2) investigating the scenario with mobile $\mathrm{TX}$ and $\mathrm{RX}$.

\appendices
\section{Proof of Theorem \ref{theorem1}}\label{app}
Using separation of variables \cite{cole2010heat}, we express $C(r,t)$ as $C(r,t)=R(r)T(t)$, where $R(r)$ and $T(t)$ are functions of $r$ and $t$, respectively. Substituting $C(r,t)$ into \eqref{fsl}, we obtain
\begin{align}\label{PDF1}
\frac{2D_\mathrm{v}R^{'}(r)}{rR(r)}+\frac{D_\mathrm{v}R^{''}(r)}{R(r)}=\frac{T^{'}(t)}{T(t)}\overset{(a)}{=}\epsilon,
\end{align}
where $R^{'}(r)=\frac{\partial R(r)}{\partial r}$, $R^{''}(r)=\frac{\partial^2R(r)}{\partial r^2}$, and $T^{'}(t)=\frac{\partial T(t)}{\partial t}$. The left-hand side of the first equality in \eqref{PDF1} is a function of $r$, denoted by $\epsilon(r)$, and the right-hand side is a function of $t$, denoted by $\epsilon(t)$. As $\epsilon(r)=\epsilon(t)$, $\epsilon(r)$ and $\epsilon(t)$ are constant $\epsilon$ that results in the equality $(a)$. Based on \eqref{PDF1}, we first obtain
\begin{align}\label{Rr}
r^2R^{''}(r)+2rR^{'}(r)-\frac{\epsilon}{D_\mathrm{v}}r^2R(r)=0.
\end{align}

By setting $\epsilon=-D_\mathrm{v}\lambda_n^2$, $n=1,2,3,...$, \eqref{Rr} becomes the Bessel function \cite{beals2016special}. For each given $\lambda_n$, the solution of $R(r)$ in \eqref{Rr}, denoted by $R_n(r)$, is given by $R_n(r)=E_nj_0(\lambda_nr)$, where $E_n$ is a constant. As $R_n(r)$ needs to satisfy the boundary condition in \eqref{bc}, we substitute $R_n(r)$ into \eqref{bc} and obtain \eqref{Dvl}. Based on \eqref{PDF1}, we then obtain $\frac{T_n^{'}(t)}{T_n(t)}=-D_\mathrm{v}\lambda_n^2.$
By considering an obvious condition $T_n(t\rightarrow\infty)=0$, one principle solution to $T_n(t)$ is $T_n(t)=B_n\exp\left(-D_\mathrm{v}\lambda_n^2t\right)$. Based on the principle of superposition, $C(r,t)$ then becomes $C(r,t)=\sum_{n=1}^{\infty}R_n(r)T_n(t)=\sum_{n=1}^{\infty}E_nB_nj_0\left(\lambda_nr\right)\exp\left(-D_\mathrm{v}\lambda_n^2t\right).$

We next determine $E_nB_n$ based on the initial condition in \eqref{IC}. According to \cite[eq. (12)]{dincc2018impulse} and \cite[eq. (21)]{arjmandi2019diffusive}, if $n\neq n^{'}$, $rj_0(\lambda_nr)$ and $rj_0(\lambda_{n^{'}}r)$ are orthogonal to each other. We then derive $\int_{0}^{r_{\ss\T}}r^2j_0(\lambda_nr)j_0(\lambda_{n^{'}}r)\mathrm{d}r$ as
\begin{align}\label{kk}
\int_{0}^{r_{\ss\T}}r^2j_0(\lambda_nr)j_0(\lambda_{n^{'}}r)\mathrm{d}r=\left\{\begin{array}{lr}
0,~~~~~~~~~~~~~~~~~\;n\neq n^{'},\\
\frac{2\lambda_nr_{\ss\T}-\mathrm{sin}\left(2\lambda_nr_{\ss\T}\right)}{4\lambda_n^3}, n=n^{'}.
\end{array}
\right.
\end{align}

We substitute $C(r,t)$ into \eqref{IC}, and then multiply $j_0\left(\lambda_nr\right)$ to both sides of the equality. With some mathematical manipulations and taking the integral of both sides with respect to $r$, we obtain
\begin{align}\label{anbn}
\sum_{n=1}^{\infty}E_nB_n\!\!\int_{0}^{r_{\ss\T}}r^2j_0^2(\lambda_nr)\mathrm{d}r=\frac{1}{4\pi}\int_{0}^{r_{\ss\T}}\delta(r)j_0(\lambda_nr)\mathrm{d}r.
\end{align}

Based on \eqref{kk} and \eqref{anbn}, we obtain $E_nB_n=\frac{\lambda_n^3}{\pi(2\lambda_nr_{\ss\T}-\mathrm{sin}(2\lambda_nr_{\ss\T}))}$, and then obtain $C(r,t)$ accordingly. Based on \cite[eq. (3.106)]{schulten2000lectures}, the vesicle fusion probability, which is also the release probability of molecule $\sigma$, is expressed as $f_\mathrm{r}(t)=4\pi r_{\ss\T}^2k_\mathrm{f}C(r_{\ss\T},t)$. By substituting $C(r_{\ss\T},t)$ into $f_\mathrm{r}(t)$, we obtain \eqref{ff}.

\section{Proof of Lemma \ref{hrb}}\label{ab}
We perform the surface integral by considering a small surface element that is the lateral surface of a conical frustum, denoted by $\mathrm{d}S$, on the $\mathrm{TX}$ membrane. The point $\alpha$ is on $\mathrm{d}S$ with the coordinates $(x, y, z)$. The base and top radii of this conical frustum are $y$ and $y+\mathrm{d}y$, respectively, and the slant height is $\sqrt{\mathrm{d}^2x+\mathrm{d}^2y}$. Based on the expression for the lateral surface area of a conical frustum, $\mathrm{d}S$ is calculated as $\mathrm{d}S=2\pi y\sqrt{1+\left(\frac{\mathrm{d}y}{\mathrm{d}x}\right)^2}\mathrm{d}x+\pi\frac{\mathrm{d}y}{\mathrm{d}x}\sqrt{1+\left(\frac{\mathrm{d}y}{\mathrm{d}x}\right)^2}\mathrm{d}^2x$. Since $y=\sqrt{r_{\ss\T}^2-x^2}$, we have $\frac{\mathrm{d}y}{\mathrm{d}x}=-\frac{x}{\sqrt{r_{\ss\T}^2-x^2}}$. By substituting $\frac{\mathrm{d}y}{\mathrm{d}x}$ into $\mathrm{d}S$, we obtain $\mathrm{d}S=2\pi r_{\ss\T}\mathrm{d}x-\frac{\pi xr_{\ss\T}}{r_{\ss\T}^2-x^2}\mathrm{d}^2x\approx 2\pi r_{\ss\T}\mathrm{d}x$, where $\frac{\pi xr_{\ss\T}}{r_{\ss\T}^2-x^2}\mathrm{d}^2x$ is omitted since it is the higher order infinitesimal. As $\mathrm{d}S$ is infinitesimal, we treat the distance between each point on $\mathrm{d}S$ and the center of the $\mathrm{RX}$ as $l_\alpha$, where $l_\alpha$ can be expressed based on $x$ as $l_\alpha=\sqrt{r_{\ss\T}^2+l^2-2lx}$. By substituting $l_\alpha$ into \eqref{pa}, we obtain $p_\alpha(t,x)$. Accordingly, the hitting probability at the $\mathrm{RX}$ for molecules released from $\mathrm{d}S$ is $\rho p_\alpha(t,x)\mathrm{d}S$. Furthermore, $p_\mathrm{u}(t)$ is obtained by integrating $\rho p_\alpha(t,x)\mathrm{d}S$, i.e., $p_\mathrm{u}(t)=\int_{\Omega_\mathrm{t}}\rho p_\alpha(t,x)\mathrm{d}S=\int_{-r_{\ss\T}}^{r_{\ss\T}}2\pi r_{\ss\T}\rho p_\alpha(t,x)\mathrm{d}x$. By substituting $p_\alpha(t,x)$ into $p_\mathrm{u}(t)$ and solving the integral, we obtain \eqref{pbt}.
\bibliographystyle{IEEEtran}
\bibliography{refsix}

\begin{thebibliography}{10}
\providecommand{\url}[1]{#1}
\csname url@samestyle\endcsname
\providecommand{\newblock}{\relax}
\providecommand{\bibinfo}[2]{#2}
\providecommand{\BIBentrySTDinterwordspacing}{\spaceskip=0pt\relax}
\providecommand{\BIBentryALTinterwordstretchfactor}{4}
\providecommand{\BIBentryALTinterwordspacing}{\spaceskip=\fontdimen2\font plus
\BIBentryALTinterwordstretchfactor\fontdimen3\font minus
  \fontdimen4\font\relax}
\providecommand{\BIBforeignlanguage}[2]{{%
\expandafter\ifx\csname l@#1\endcsname\relax
\typeout{** WARNING: IEEEtran.bst: No hyphenation pattern has been}%
\typeout{** loaded for the language `#1'. Using the pattern for}%
\typeout{** the default language instead.}%
\else
\language=\csname l@#1\endcsname
\fi
#2}}
\providecommand{\BIBdecl}{\relax}
\BIBdecl

\bibitem{farsad2016comprehensive}
N.~Farsad, H.~B. Yilmaz, A.~Eckford, C.-B. Chae, and W.~Guo, ``A comprehensive
  survey of recent advancements in molecular communication,'' \emph{IEEE
  Commun, Surveys Tuts.}, vol.~18, no.~3, pp. 1887--1919, 3rd Quarter, 2016.

\bibitem{ahmadzadeh2016comprehensive}
A.~Ahmadzadeh, H.~Arjmandi, A.~Burkovski, and R.~Schober, ``Comprehensive
  reactive receiver modeling for diffusive molecular communication systems:
  Reversible binding, molecule degradation, and finite number of receptors,''
  \emph{IEEE Trans. Nanobiosci.}, vol.~15, no.~7, pp. 713--727, Oct. 2016.

\bibitem{jamali2019channel}
V.~Jamali, A.~Ahmadzadeh, W.~Wicke, A.~Noel, and R.~Schober, ``Channel modeling
  for diffusive molecular communication--a tutorial review,'' \emph{Proc.
  IEEE}, Jun. 2019.

\bibitem{yilmaz2017chemical}
H.~B. Yilmaz, G.-Y. Suk, and C.-B. Chae, ``Chemical propagation pattern for
  molecular communications,'' \emph{IEEE Wireless Commun. Lett.}, vol.~6,
  no.~2, pp. 226--229, Apr. 2017.

\bibitem{arjmandi2016ion}
H.~Arjmandi, A.~Ahmadzadeh, R.~Schober, and M.~N. Kenari, ``Ion channel based
  bio-synthetic modulator for diffusive molecular communication,'' \emph{IEEE
  Trans. Nanobiosci.}, vol.~15, no.~5, pp. 418--432, Jul. 2016.

\bibitem{schafer2019spherical}
M.~Sch{\"a}fer, W.~Wicke, W.~Haselmayr, R.~Rabenstein, and R.~Schober,
  ``Spherical diffusion model with semi-permeable boundary: A transfer function
  approach,'' in \emph{Proc. IEEE ICC}, Jun. 2020, pp. 1--7.

\bibitem{jahn1999membrane}
R.~Jahn and T.~C. S{\"u}dhof, ``Membrane fusion and exocytosis,'' \emph{Annu.
  Rev. Biochem.}, vol.~68, no.~1, pp. 863--911, Jul. 1999.

\bibitem{biomor}
A.~Morgan, ``Exocytosis,'' \emph{Essays BioChem.}, vol.~30, pp. 77--95, 1995.

\bibitem{bonifacino2004mechanisms}
J.~S. Bonifacino and B.~S. Glick, ``The mechanisms of vesicle budding and
  fusion,'' \emph{Cell}, vol. 116, no.~2, pp. 153--166, Jan. 2004.

\bibitem{kyoung2008vesicle}
M.~Kyoung and E.~D. Sheets, ``Vesicle diffusion close to a membrane:
  intermembrane interactions measured with fluorescence correlation
  spectroscopy,'' \emph{Biophys. J.}, vol.~95, no.~12, pp. 5789--5797, Dec.
  2008.

\bibitem{hay2007calcium}
J.~C. Hay, ``Calcium: a fundamental regulator of intracellular membrane
  fusion?'' \emph{EMBO Rep.}, vol.~8, no.~3, pp. 236--240, Mar. 2007.

\bibitem{arjmandi2019diffusive}
H.~Arjmandi, M.~Zoofaghari, and A.~Noel, ``Diffusive molecular communication in
  a biological spherical environment with partially absorbing boundary,''
  \emph{IEEE Trans. Commun.}, vol.~67, no.~10, pp. 6858--6867, Jul. 2019.

\bibitem{erban2007reactive}
R.~Erban and S.~J. Chapman, ``Reactive boundary conditions for stochastic
  simulations of reaction--diffusion processes,'' \emph{Physical Biology},
  vol.~4, no.~1, p.~16, 2007.

\bibitem{chang2005physical}
R.~Chang, \emph{Physical {Chemistry} for the {Biosciences}}.\hskip 1em plus
  0.5em minus 0.4em\relax Sausalito, CA, USA: Univ. Science Books, 2005.

\bibitem{dincc2018impulse}
F.~Din{\c{c}}, B.~C. Akdeniz, A.~E. Pusane, and T.~Tugcu, ``Impulse response of
  the molecular diffusion channel with a spherical absorbing receiver and a
  spherical reflective boundary,'' \emph{IEEE Trans. Mol. Biol. Multi-Scale
  Commun.}, vol.~4, no.~2, pp. 118--122, Jun. 2018.

\bibitem{berg1993random}
H.~C. Berg, \emph{Random {Walks} in {Biology}}.\hskip 1em plus 0.5em minus
  0.4em\relax Princeton, NJ, USA: Princeton Univ. Press, 1993.

\bibitem{crank1979mathematics}
J.~Crank, \emph{The {Mathematics} of {Diffusion}}.\hskip 1em plus 0.5em minus
  0.4em\relax London, U.K.: Oxford Univ. Press, 1979.

\bibitem{olver1960bessel}
F.~W. Olver and L.~C. Maximon, \emph{Bessel {Functions}}.\hskip 1em plus 0.5em
  minus 0.4em\relax New York, NY: Cambridge Univ. Press, 1960.

\bibitem{heren2015effect}
A.~C. Heren, H.~B. Yilmaz, C.-B. Chae, and T.~Tugcu, ``Effect of degradation in
  molecular communication: Impairment or enhancement?'' \emph{IEEE Trans. Mol.
  Biol. Multi-Scale Commun.}, vol.~1, no.~2, pp. 217--229, Jun. 2015.

\bibitem{noel2017simulating}
A.~Noel, K.~C. Cheung, R.~Schober, D.~Makrakis, and A.~Hafid, ``Simulating with
  {AcCoRD}: Actor-based communication via reaction--diffusion,'' \emph{Nano
  Commun. Netw.}, vol.~11, pp. 44--75, Mar. 2017.

\bibitem{cole2010heat}
K.~Cole, J.~Beck, A.~Haji-Sheikh, and B.~Litkouhi, \emph{Heat {Conduction}
  {Using} {Greens} {Functions}}.\hskip 1em plus 0.5em minus 0.4em\relax Boca
  Raton, FL, USA: CRC Press, 2010.

\bibitem{beals2016special}
R.~Beals and R.~Wong, \emph{Special {Functions} and {Orthogonal}
  {Polynomials}}.\hskip 1em plus 0.5em minus 0.4em\relax Cambridge, U.K.:
  Cambridge Univ. Press, 2016, vol. 153.

\bibitem{schulten2000lectures}
K.~Schulten and I.~Kosztin, \emph{Lectures in {Theoretical}
  {Biophysics}}.\hskip 1em plus 0.5em minus 0.4em\relax Univ. Illinois,
  Champaign, IL, USA, 2000, vol. 117.

\end{thebibliography}
\end{document}